# Optically Induced Transparency in a Micro-cavity

**Yuanlin Zheng[1], Jianfan Yang[2], Zhenhua Shen[1], Jianjun Cao[1], Xianfeng Chen[1*], Xiaogan Liang[3], and Wenjie Wan[1,2*]**

[1]MOE Key Laboratory for Laser Plasmas and Collaborative Innovation Center of IFSA, Department of Physics and Astronomy, Shanghai Jiao Tong University, Shanghai 200240, China
[2]The University of Michigan-Shanghai Jiao Tong University Joint Institute, The State Key Laboratory of Advanced Optical Communication Systems and Networks, Shanghai Jiao Tong University, Shanghai 200240, China
[3]Department of Mechanical Engineering, University of Michigan, Ann Arbor, MI 48109, USA
*Correspondence and requests for materials should be addressed to:
Wenjie Wan (wenjie.wan@sjtu.edu.cn) or Xianfeng Chen (xfchen@sjtu.edu.cn)

## Abstract

Electromagnetically induced transparency has the unique ability to optically control transparency windows with low light in atomic systems. However, its practical applications in quantum physics and information science are limited due to rigid experimental requirements. Here, we demonstrate a new mechanism of optically induced transparency in a micro-cavity by introducing a four-wave mixing gain to nonlinearly couple two separated resonances of the micro-cavity in an ambient environment. A signature Fano-like resonance was observed owing to the nonlinear interference of the two coupled resonances. Moreover, we show that the unidirectional gain of the four-wave mixing can lead to the remarkable effect of non-reciprocal transmission at the transparency windows. Optically induced transparency may offer a unique platform for a compact, integrated solution to all-optical and quantum information.

## Introduction

Electromagnetically induced transparency (EIT) is one of the greatest discoveries to verify the quantum interference nature of atomic systems [1]. It allows the flipping of an optical opaque transmission window to a transparent one under a secondary coherent illumination beam even at the few-photon limit[2-3], which immediately provides an opportunity for light-matter-light interactions that are in great demand by all-optical and quantum information processing[4]. However, the long-lasting challenge is that direct interactions between photons are prohibited unlike electrons in an electronic system. EIT was first proposed and observed in atomic gases [1], and has also been observed in solid state systems [5, 6]. Much attention has been drawn to its unique properties, including slowing or even stopping light, all-optical switching, and photon storage [6-9]. This provides tremendous opportunities for quantum physics and information science. However, in most circumstances, those atoms must be prepared in gas phase, which requires optical cooling or vapor heating techniques combined with vacuum isolation, making it difficult for chip-scale integrations, which are the major proposed applications. Alternatively, classical analogies that mimic the EIT effect are under active pursuit in various physical systems, including coupled resonators [10, 11], photonic crystals [12,13], and plasmonic meta-materials [14]. However, most of these rely on linear coupling between the resonances, which can only exhibit EIT-like spectra but lack of actively controlled transparency. Until recently, successful attempts [15-18] have cloned the idea of EIT in an opto-mechanical micro-cavity to induce a narrow transparency with the aid of mechanical oscillations that are excited through Brillouin scattering (BS) nonlinearity, which provide enough phonons to couple some hybrid optical-mechanical resonances that are similar to their counterparts, photons in the EIT. However, most of these attempts still require cooling and vacuum systems to preserve the high-quality mechanical modes; furthermore, careful designs must be addressed for both the mechanical and optical modes.

In this work, we were motivated by the EIT in atomic systems. We report an optically induced transparency (OIT) scheme in a compact micro-cavity in an ambient environment by exploring cavity-enhanced four-wave mixing (FWM) gain to introduce a transparency window in an opaque resonance dip. This resonance dip directly results from the interference between two resonances that are coupled nonlinearly through the FWM process. Active controlling of the OIT can be achieved by varying a strong pump beam, and small frequency detuning of the pump can lead to a Fano-like asymmetric resonance to justify the interference nature of the OIT. Furthermore, the OIT observed here is a non-reciprocal one because the FWM gain is unidirectional owing to the conservation law of momentum. Our unique OIT scheme offers a completely new platform to study analogical atomic quantum interference effects in a simple manner. Moreover, it can be further exploited for critical on-chip photonics applications, such as optical isolator, all-optical switching, and wavelength conversion.

## Materials and Methods

We developed an OIT window inside a single optical resonance dip of a whispering gallery mode (WGM) micro-cavity through a cavity-enhanced FWM scheme, as shown in Fig. 1. In the waveguide coupled micro-cavity, light can tunnel from the input waveguide into the micro-cavity to form the whispering gallery resonance mode by total internal reflection [19]. The resonance conditions lead to opaque dips in the transmission spectrum, as shown in Fig. 1b, that is even perfectly opaque in critical coupling conditions [20, 21]. Traditionally, such resonance line-shapes cannot be modified unless the intrinsic properties of the cavity are altered, e.g., size, surface scattering, or absorption loss [19]. However, to obtain a transparency in an opaque resonance dip, i.e., an EIT-like transmission spectrum, an external narrow gain/loss mechanism must be introduced. In atomics physics, this can be achieved by atomic resonant enhanced processes; however, in optics, cavity resonances facilitate the exploration of this narrow transparency, e.g., using two-coupled resonators with adjacent resonances [10-12], double-layered resonators[22] with close resonances, or BS gain in an opto-mechanical micro-cavity [15-18]. However, these methods lack active all-optical control of the induced transparency. Furthermore, a hybrid approach, which combines ultra-slow light atomic resonance with photonic structured resonance, can also lead to induced transparency [13]. Again, however, this may still require cooling, vacuum systems, and even a trapping mechanism to place the atoms within the photonic structure to maintain the atomic resonances.

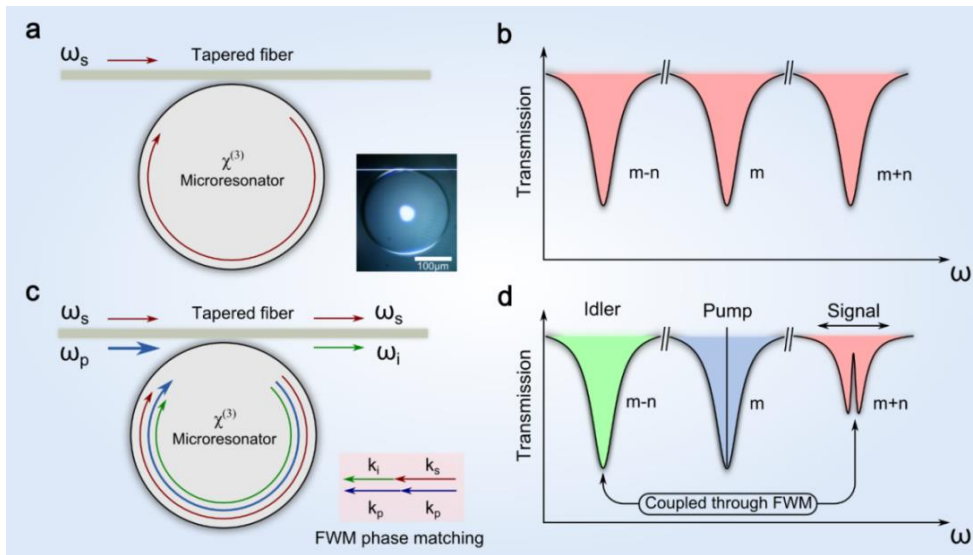

Figure 1 | Schematic illustration of the optically induced transparency. (a) Light coupling into a micro-cavity through a tapered fiber at near-critical coupling conditions, providing (b) a transmission spectrum of the signal wave, $\omega_s$. (c)

Introducing a strong pump wave, $\omega_p$, into the same micro-cavity and generating a nonlinear FWM by fulfilling the phase matching condition. (d) A transparent window inside the transmission dip can be induced due to the nonlinear coupling of the two resonances at the signal and at the idler waves via the FWM process.

Here, we introduce a narrow parametric gain from a nonlinear degenerated FWM process to open up a transparency inside the opaque resonance dip. To achieve this, a strong pump beam is injected into one resonance at a frequency of $\omega_m$ (m is the azimuthal order of WGM modes) of a micro-cavity with $\chi^{(3)}$ nonlinearity, as shown in Fig. 1c. This enables degenerated FWMs to be enhanced by the cavity resonances, providing parametric gains to two other distanced resonances that are symmetrically located at frequencies $\omega_{m+n}$ and $\omega_{m-n}$, which keeps $\omega_{m+n} + \omega_{m-n} = 2\omega_m$. Meanwhile, a second signal beam scans through the resonance near $\omega_{m+n}$, and a transparency peak can be obtained inside the original resonance dip, as shown in Fig. 1d. The parametric gain is a narrow-linewidth one owing to the cavity resonance's enhancement; hence, its linewidth is imprinted by the resonances'. As explained below, for a given narrow linewidth (compared to the resonance's linewidth) pump, the transparency window inside the opaque dip of the signal transmission will copy the linewidth of the resonance near the idler wave's frequency, $\omega_{m-n}$. Effectively, two resonances near the signal and idler waves' frequencies are nonlinearly coupled together via the FWM process. They interfere with each other, resulting in a hump-in-dip transmission spectrum. Note that, a broad-spectrum gain/loss like scattering loss and absorption cannot lead to this OIT transmission, and it only affects the quality factor (Q-factor) of the resonance [19]. Unlike the previous works, e.g., opto-mechanically induced transparency (OMIT) or Brillion scattering (BS) induced transparency (BSIT) [15-18], in which the narrow acoustic induced gain requires careful tailoring both in the mechanical and optical modes to fulfill rigid energy and momentum conservations, our scheme does not require manipulating the two species of resonance modes and does not demand additional cooling or vacuum systems to stabilize those resonances. Hence, it can be ubiquitously attained in many resonant systems with $\chi^{(3)}$ nonlinearity.

To gain more physical insights, we now provide a precise discussion of the OIT theorem. For FWM in a resonant cavity system, its signal, $A_s(\omega_s)$, and idler, $A_i(\omega_i)$, beams can be coupled in a nonlinear fashion, which can be formulated in a set of coupled equations according to the coupled mode theory, as shown in Eqs. 1 & 2 (see supplement). These cavity resonances are governed by the photon decay rate, $\kappa$ ($\kappa = 1/\tau$, $\tau$ is the photon lifetime), including the internal rate, $\kappa_o$ (due to internal absorption, scattering and radiation loss), and the external rate, $\kappa_e$ (due to waveguide coupling), with a frequency detuning of $\Delta\omega$ from its central resonance of $\omega_o$ for both the signal, $\omega_s$, and the idler, $\omega_i$ (s and i represent the signal and idler waves, respectively). The two frequency detuning factors, $\Delta\omega_s$ and $\Delta\omega_i$, depend on each other through the FWM energy conservation: $\omega_s + \omega_i = 2\omega_p$, and they are on the scale of a single cavity resonance's linewidth ($\sim 1/\tau$). The most important coupling term, g, arises from nonlinear FWMs, effectively linking the two resonances of the signal and the idler together. In two-coupled micro-cavities [10-12], the linear coupling initiates EIT-like phenomena by interfering with two adjacent resonance modes. While our approach of utilizing FWM to nonlinearly tie two separated resonances is genuinely different, only relevant works can be found on EIT from a multiple resonant atomic system [23]. Even compared to FWM-based EIT in an atomic system, our OIT is achieved on a degenerated FWM scheme, and it only requires two inputs near 1550 nm telecommunication windows, which is very difficult to access with atomic systems.

$$\frac{dA_s(\omega_s)}{dt} = (-\kappa_{os} - \kappa_{es} + i\Delta\omega_s)A_s(\omega_s) + i\sqrt{2\kappa_{es}}A_s^{in}(\omega_s) + igA_i^*(\omega_i) \tag{1}$$

$$\frac{dA_i(\omega_i)}{dt} = (-\kappa_{oi} - \kappa_{ei} + i\Delta\omega_i)A_i(\omega_i) + igA_s^*(\omega_s) \tag{2}$$

Under the assumption of FWM phase matching conditions (Fig. 1c), which can be obtained in a micro-cavity with small dispersion, the transmission rate of the signal (normalized with input) reads as:

$$T_s = (1 + 2\frac{\kappa_{es}X_i^*}{X_sX_i^* - g^2})^2, \tag{3}$$

where $X = -\kappa_o - \kappa_e + i\Delta\omega$ (s and i represent signal and idler waves, respectively). The transmission $T_s$ is affected by both resonances of the signal and idler. Moreover, it can be controlled by the nonlinear coupling term, g, which originated from the third-order Kerr nonlinearity [see supplement]. Because both the signal and idler waves are confined in the same micro-cavity and their wavelengths are close (few nm), the decay rates of both resonances can be assumed to be the same. Furthermore, under the zero detuning condition of $\Delta\omega_s = -\Delta\omega_i$, a transparency window can be observed in the $T_s$ with proper g factors, as shown in Fig. 2a, which provides the transmission spectra of the signal and the idler under different coupling strengths of g. As the nonlinear coupling strength gradually increases, the original opaque resonance dip first becomes shallow, then transits into a hump-in-dip spectrum and finally becomes an amplified peak above unity.

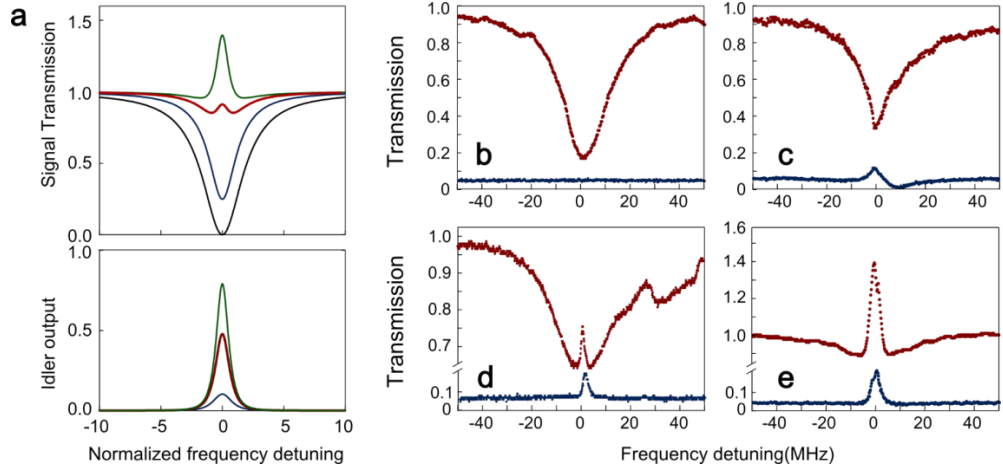

Figure 2 | Theoretical and experimental transmission spectra of the signal and the idler under various nonlinear gain strengths for the zero detuning condition of $\Delta\omega_s = -\Delta\omega_i$. (a) The theoretical results of the signal's transmission (upper) and the idler's generation coordinate (lower) with an increasing gain factor of $g = 0, 0.3, 0.41, 0.45$ for the black, blue, red and green curves. (b)- (d) Experimental measurements with increasing pump intensities of 0 mW, 2 mW, 3 mW, and 5 mW, respectively. The OIT manifests itself in (d).

## Results and Discussion

The OIT arises from the interference of the signal and the idler resonances through nonlinear coupling during FWMs in which the gain induced near the signal resonance is a narrow one enhanced by two other resonances at the pump and the idler. For a narrow linewidth pump (launched at one resonance of the micro-cavity), its FWM gain near the signal resonance purely imprints the idler resonance's linewidth through the nonlinear coupling, g, in Eqs. 1 and 2 to effectively create a virtual resonance peak near the signal resonance. This hump-in-dip spectrum near the signal resonance in Fig. 2a results from the interference between the virtual gain peak and the original resonance dip. Meanwhile, a narrow gain peak, which grows with increasing g, is initiated simultaneously at the idler resonance in Fig. 2a, which verifies this cross-coupling. Unlike the EIT case, the dispersion relation during OIT is not significantly distorted because the widths of both resonances are relatively broad compared to the atomic resonance's linewidth in EIT or the BS gain's in OMIT. Hence, the slow light effect here can only be multiplied a few times (see supplement), and the fast light effect is also absent in OIT. However, this may offer a unique opportunity for ultrafast optical switching applications, which requires a fast photon decay rate, i.e., broad resonance [24].

Experimentally, we verify the above theory by a tapered fiber-coupled silica microsphere cavity, as shown in Fig. 1, whose whispering gallery modes can be accessed by 1550 nm telecommunication lasers for both the pump and signal waves. The size of the microsphere is approximately 250 μm, and it is carefully designed with weak anomalous dispersion to encourage nonlinear FWMs. A strong pump beam is injected into one of the resonances at approximately 1548.52 nm, and it is further locked to this resonance through a thermal self-locking mechanism [25]. When pumping above a threshold (several mW), a stable optical frequency comb [26, 27] can be generated through FWMs, which offers us a reference to scan a second signal beam at one of the comb lines at approximately 1535.64 nm. Afterwards, we decrease the pump power under the threshold of the optical comb generation to observe the OIT. Fig. 2 b-e show the experimental observations of the OIT spectra of the signal wave and their corresponding idler spectra under various pump intensities. Without the pump, the signal wave approaches a near-critical coupling situation, allowing only less than 10% transmission at the center of the resonance. The linewidth of the resonance dip is approximately 20 MHz, which corresponds to a Q-factor of approximately $10^7$. After the injection of a weak pump at approximately 2 mW, the spectrum opens up with an approximate 30% transmission at the dip, and the linewidth shrinks to approximately 15 MHz. Here, the third-order nonlinear susceptibility, $\chi^{(3)}$, of silica is $2.5 \times 10^{-22}$ m$^2$/V$^2$, which corresponds to $g \approx 0.3$ in the numerical simulation [Fig. 2a, see supplement]. When the pump intensity boosts to approximately 3 mW, a signature hump begins to exhibit itself inside the dip in Fig. 2d. This hump spans approximately 5 MHz, which is broader than the transparency windows in EIT or OMIT. Finally, after the pump intensity reaches the threshold, an amplified peak replaces the original dip in Fig. 2e. Additionally, the idler waves' intensity grows coordinately with the increasing pumping. Moreover, if pumping harder, this process can be cascaded further to generate multiple amplified peaks at other resonances nearby to form an optical comb [27]. Notably, a small transmission jump manifests in the OIT spectra (Fig. 2d), and this may be caused by interference with an adjacent higher order resonance mode to produce a Fano-like spectrum[22], which we discuss in the next section. Additionally, the linewidth of OIT can be approximated by $\Gamma = \kappa \frac{g^2 - \kappa^2}{g^2 + \kappa^2}$ for $g > \kappa$ (above the OIT threshold, where κ is the total decay rate for both signal & idler resonances) or by $\Gamma = \kappa \frac{\kappa^2 - g^2}{\kappa^2 + g^2}$ for $g < \kappa$, (below the OIT threshold). This result explains the linewidth narrowing effect in Fig. 2c, and the sharp OIT peak in Fig. 2d [see supplement].

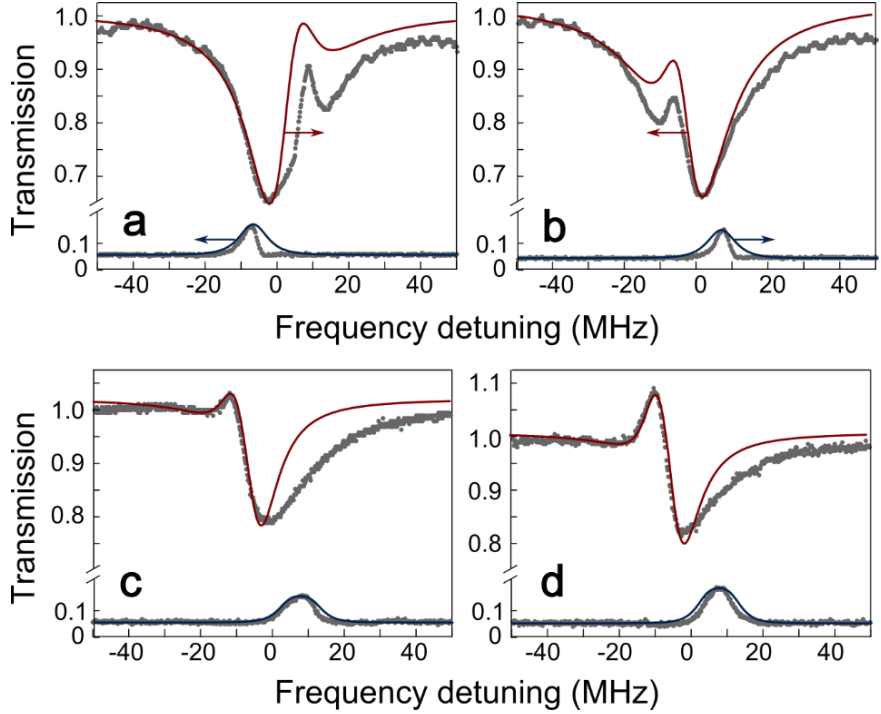

Figure 3 | Fano-like resonance spectra for a non-zero detuning condition. (a), (b) asymmetric Fano-like resonance spectra by blue and red-detuning of the pump. (c) Fano-like spectrum by further red-detuning of the pump by increasing the gain in (d). Dotted lines are measured experimental results, and the solid lines correspond to the calculations that are based on Eq. 3. Upper curve: signal beam, lower curve: idler beam.

**Fano-like resonance in OIT**

Interestingly, an asymmetric resonance in Fig. 3 can be attained via the small detuning of two coupled resonances. This can be recognized as the Fano-Feshbach resonance [28, 29], which is known to originate from the interference of two neighboring resonances and provides an asymmetric profile of resonances. Here, we can create a non-zero detuning scenario of $\Delta\omega_s \neq -\Delta\omega_i$, by either (1) setting $\omega_{so} + \omega_{io} \neq 2\omega_{po}$, which could be found for a non-phase-matched resonator, or by (2) detuning the frequency of the pump off its resonance center, $\omega_{po}$, in a phase-matched scheme (see supplement). Experimentally, we adapted the latter approach by slightly detuning the pump's frequency around the central resonance. Because the relationship of $\omega_s + \omega_i = 2\omega_p$ and the phase matching condition must be maintained, this effectively shifts the virtual resonance peak off the center of the signal resonance while pushing the idler gain peak to the other direction, as shown in Fig. 3. Ideally, the Fano resonance enables a constructive interference on one side and a destructive one on the other side due to a π phase jump at the center of one resonance [28, 29]. This is clearly pronounced by comparing the two OIT spectra in Fig. 3a, b when blue-detuning or red-detuning the pump at the location where the Fano-like resonance's symmetry flips. Further red-detuning the pump pushes the OIT peak off to the edge, and it can be amplified by increasing the pump intensity. Attaining a Fano-like resonance in our case is allowed due to the similar linewidths of the two coupled resonances; however, in other situations in which the two resonances' linewidths significantly differ from each other, e.g., OMIT or BSIT, such phenomenon is difficult to observe.

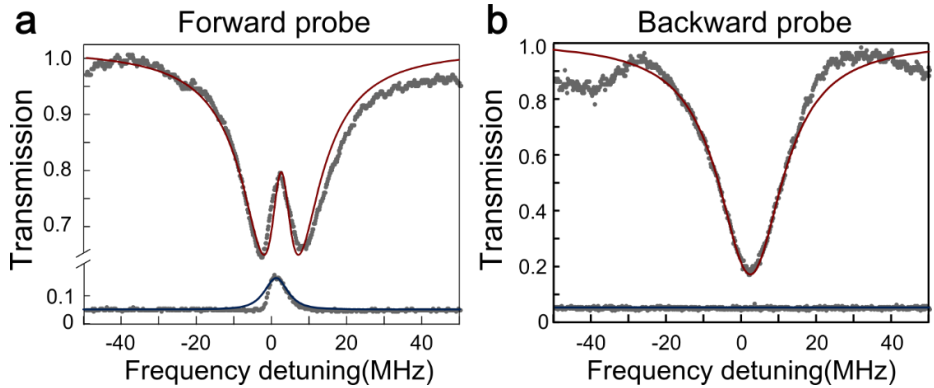

Figure 4 | Demonstration of the non-reciprocal OIT. (a) The OIT is observed in the forward direction due to the unidirectional FWM gain. (b) The transparency window is absent in the backward direction. Dotted lines are the measured experimental results, and the solid lines correspond to the calculations that are based on Eq. 3. Upper curve: signal beam, lower curve: idler beam.

**Non-reciprocal transparency in OIT**

Finally, the observed OIT here is a non-reciprocal one, as shown in Fig. 4. Clearly, the forward transmission is featured with the OIT window, which is absent in the backward transmission. The non-reciprocal distinction ratio can reach ~6.5 dB at the center of the OIT window, which can be enlarged further to approximately 26 dB under the critical coupling situation in which the backward transmission is minimized [20]. Such phenomena occurs because of the unidirectional gain nature of the FWM, which strictly follows the phase matching conditions in which the forward momentum carried by the pump only enables the FWM gain in the same direction while inhibiting the gain in reverse. In the microwave regime, a similar mechanism has been achieved in a non-reciprocal solid state superconducting qubits device [30]. However, in optics, a BSIT utilizing a non-reciprocal BS gain has recently been demonstrated. Our novel approach utilizes unidirectional FWM gain in a micro-cavity, which features a non-reciprocal gain arising from the strong pump that has no demand for the signal's input power (below 80 μW). Hence, our method may be favorable for solving the long-standing problem of non-magnetic optical isolation in on-chip photonic applications.

## Conclusion

We have theoretically and experimentally demonstrated a new scheme of an optically induced transparency in a compact micro-cavity in an ambient environment. Because the OIT relies only on the nonlinearity and resonance properties of the medium, similar features should be ubiquitously expected among any other physical system with these characteristics. This system may offer a new avenue for a compact, integrated solution for all-optical processing and quantum information.

**Contributions:**
W.W. and Y.Z. designed the study; Y.Z. designed the experimental setup, performed the research, and analyzed the data; W.W. and X.C. supervised the work; J.Y., Z.S., and J.C. helped with the fabrication; X.L. provided advice and helpful discussion; and Y.Z. and W.W wrote the paper. All authors reviewed the manuscript.

**Acknowledgements:**
This work was supported by the National Natural Science Foundation of China (grant nos 11304201 and 61475100), the National 1000-plan Program (Youth), the Shanghai Pujiang Talent Program (grant no. 12PJ1404700), and the Shanghai Scientific Innovation Program (grant no. 14JC1402900).